# Fabrication of Asymmetric Electrode Pairs with Nanometer Separation Made of Two Distinct Metals

*Mandar M. Deshmukh, Amy L. Prieto, Qian Gu, and Hongkun Park*[*]

Department of Chemistry and Chemical Biology, Harvard University, Cambridge MA 02138, USA

We report a simple and reproducible method to fabricate two metallic electrodes made of different metals with a nanometer-sized gap. These electrodes are fabricated by defining a pair of gold electrodes lithographically and electrodepositing a second metal onto one of them. The method enables the fabrication of pairs of metallic electrodes that exhibit distinct magnetic properties or work functions. The utility of this technique is demonstrated by making single-electron tunneling devices incorporating 2-nm gold nanocrystals.

Over the last decade, electrical characterization of individual chemical nanostructures, such as molecules,[1-6] nanocrystals,[7,8] and carbon nanotubes,[9-13] has been the focus of extensive research efforts. Transport studies of devices incorporating a single nanostructure allow fundamental investigations of the interplay between the electronic motion and the internal degrees of freedom of the nanostructures. The ability to make these electrical devices is technologically relevant because it may allow the fabrication of new types of devices whose functions are defined by the chemical identity of the components.

One of the major challenges in realizing nanoscale electronic devices is the reliable fabrication of metallic electrodes with a nanometer-sized separation that enable the electrical contact to individual chemical nanostructures. To date, many different approaches have been developed to fabricate electrodes with a nanometer-sized gap, such as two-step electron beam lithography,[14] shadow evaporation,[15,16] nanopore-based methods,[17] electrodeposition,[18-20] mechanical break junction,[6,21] and electromigration-induced break junction.[4,22] Most of these techniques[4,15-20,22] have been limited to fabricating symmetric tunnel junctions where the two electrodes are made of the same metal.

In this letter, we report a simple and reproducible method to fabricate two metallic electrodes made of dissimilar metals with a nanometer-sized gap between them. The fabrication is achieved first by defining a pair of gold electrodes using electron beam lithography and then by electrodepositing a second metal onto one of the electrodes.[18-20] This method enables the fabrication of asymmetric tunnel junctions made of two different metals that exhibit distinct magnetic properties or work functions,[14] and offers important advantages over the previous fabrication methods. Most importantly, the method described here allows independent tailoring of the electrodes' properties, and therefore can be used to fabricate single-nanostructure devices with spintronic[23] and light-emitting[24] functionalities. The present method combined with current-induced junction breakage also enables the simultaneous fabrication of multiple electrode pairs in a self-limiting manner. We demonstrate the potential of our technique by depositing 2-nm nanoparticles between two electrodes and realizing single electron tunneling devices that exhibit Coulomb blockade behavior at T = 4.2 K.

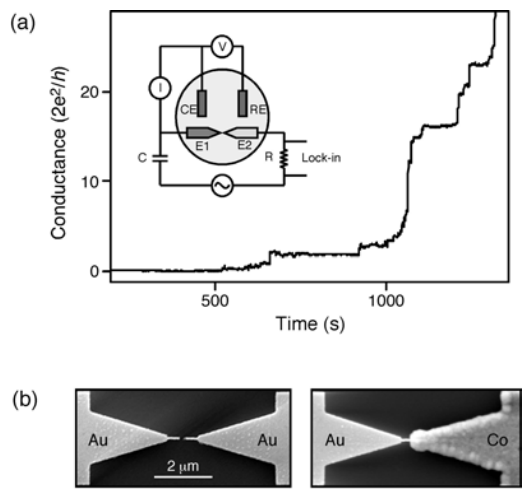

**Figure 1.** (a) Plot of the conductance between the two electrodes as a function of time during the electrodeposition of cobalt on one of the electrodes. The current used during this deposition is 7 μA. The inset shows a schematic diagram for an electrodeposition setup that allows the asymmetric metal deposition and the *in situ* resistance-monitoring. Here, RE is the reference electrode, CE is a platinum counter electrode, E1 is one of the gold electrodes onto which the metal is electrodeposited, E2 is the second electrode, C is a capacitor (440 μF), and R is a resistor (5 kΩ) across which the voltage, proportional to the detection current, is measured. (b) Scanning electron micrographs of a pair of electrodes before and after the cobalt deposition.

The technique described here builds upon electrodeposition-based methods reported previously.[18-20] The first step of the fabrication procedure is to define two gold electrodes with an initial separation of ~250 nm using conventional electron-beam lithography. The second metal is deposited subsequently onto one of the gold electrodes using solution-phase electrodeposition. In this letter, copper and cobalt are deposited to demonstrate the technique, but the same method can be extended to deposit other materials as well. The inset in Fig. 1(a) shows a schematic diagram of an electrodeposition setup that allows both asymmetric metal deposition and *in situ* resistance-monitoring. The setup consists of a potentiostat (Epsilon, Bioanalytical Systems, Inc.), a Ag/AgCl reference electrode (RE), a platinum mesh counter electrode (CE), and one gold electrode (E1) that serves as a working electrode onto which the metal is deposited. To prevent electrodeposition on the other gold electrode (E2) while allowing the resistance measurements, E1 and E2 are connected by a capacitor (C ~ 440 μF). This connection ensures that the DC voltage drops across the capacitor and that the potential of E2 floats at the electrolyte potential, preventing electrodeposition.

The electrodeposition of cobalt or copper on E1 was carried out by flowing a constant DC current between E1 and CE while monitoring the voltage drop between E1 and RE. Cobalt was deposited from an aqueous solution of 0.22 M cobalt sulfate, 0.2 M citric acid, and 0.12 M potassium citrate. Copper was deposited from an aqueous solution of 0.02 M copper sulfate, and 0.2 M potassium sulfate. The concentrations and pH of the solutions were optimized in order to promote the deposition of smooth, continuous films, and the deposition rate was tuned by controlling the current. Generally, slower depositions result in smoother, denser metallic films.

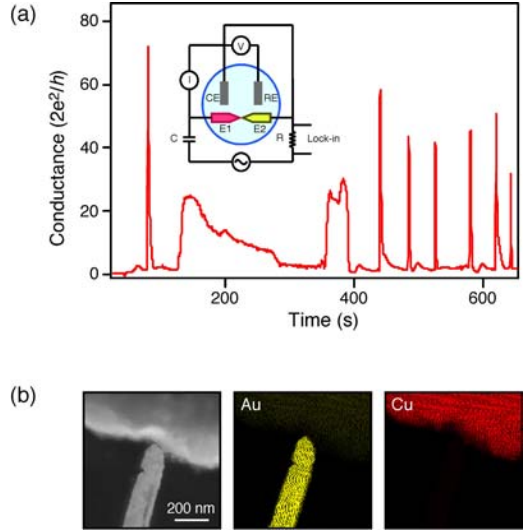

**Figure 2.** Figure caption. (a) Plot of the conductance between the two electrodes as a function of time during the electrodeposition of copper on one of the electrodes using a self-limiting electrodeposition method. The inset shows a schematic diagram for the corresponding deposition setup. (b) Leftmost panel: dark-field transmission electron micrograph of a pair of electrodes defined on a 50-nm thick silicon nitride membrane. Middle and rightmost panels: energy-dispersive X-ray spectroscopy images of the same electrode pair that provide the elemental maps for gold and copper, respectively.

The resistance between E1 and E2 was monitored by measuring the AC voltage drop across the resistor R (5 kΩ) using a lock-in system.[18-20] The frequency used in the lock-in detection was 7.9 Hz, and it was chosen to minimize the capacitive impedance while preserving sufficient measurement sensitivity. Figure 1(a) shows the conductance measured between E1 and E2 as a function of time during the deposition of cobalt. The discrete jumps on the order of one quantum of conductance $2e^2/h$ (the value of $e^2/h$ is 38.8 μS or $(25.8 kΩ)^{-1}$) can be seen clearly in Fig. 1(a), indicating that the joining of the two electrodes happens gradually a few atoms at a time. Once E1 and E2 are joined, the metal deposition starts to occur on E2. In order to prevent this unwanted deposition, the current flow was stopped immediately after the first conductance jump was observed. Figure 1 (b) shows scanning electron micrographs of a electrode pair before and after the deposition of cobalt.

The deposition setup shown in Fig. 1(a) can be modified so that the distance between the two electrodes is kept at a few nanometers in a self-limiting fashion irrespective of the initial separation. This modification involves connecting CE to the gold electrode E2, which forces the electrochemical potentials of E2 and CE to be identical. In this setup, the large current, which would normally flow through the electrolyte, passes through the small metal bridge connecting E1 and E2 once the two electrodes touch each other. The current induces the electromigration[22] or dissolution of atoms forming the metal bridge, leading to its breakage. Any metal atoms deposited on E2 are oxidized and dissolved away during the period at which E1 and E2 are disconnected, thereby ensuring that no unwanted deposition occurs on E2. Figure 2(a) shows a representative conductance trace observed during this joining and breaking process. The conductance signal shows a series of spikes signaling the repeated joining and breaking of the two electrodes during the electrodeposition process.

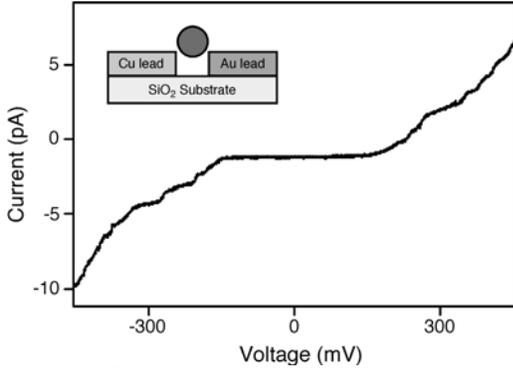

**Figure 3** Current versus voltage curve obtained at T = 4.2 K from a single electron tunneling device with a 2-nm gold nanocrystal between gold and copper electrodes.

Figure 2(b) illustrates transmission electron microscope images of a pair of electrodes fabricated using the self-limiting deposition process and shows that the two electrodes are indeed composed of distinct metals. These electrodes are made first by fabricating a pair of gold electrodes on a 50-nm thick silicon nitride membrane,[25] and subsequently electrodepositing copper onto one of the gold electrodes. The electrodes were imaged using a VG HB603 scanning transmission electron microscope. The leftmost panel in Fig. 2(b) shows a dark field image of a pair of electrodes, and the middle and rightmost panels show energy-dispersive X-ray spectroscopy images that provide the elemental maps for gold and copper, respectively. The images clearly show the asymmetric nature of the electrodes and the abruptness of the interface.

The self-limiting electrodeposition process illustrated in Fig. 2(a) allows the parallel fabrication of multiple tunnel junctions, and hence offers a critical advantage over the fabrication methods reported previously.[4,14-20,22] In a typical lithographic process, such as those employed to fabricate the initial gold electrodes at the left panel of Fig. 1(b), the separation between the two electrodes cannot be controlled with nanometer accuracy due to subtle variations in fabrication conditions. The variation in the electrode separation translates into a different electrodeposition duration for each pair of electrodes, thereby prohibiting the parallel fabrication of multiple electrode pairs in a single electrodeposition step. The self-limiting nature of the junction formation demonstrated in Fig. 2(a) minimizes the effect of the initial template separation, and allows the parallel fabrication of several electrode pairs with nanometer-sized gap.

Asymmetric tunnel junctions fabricated by electrodeposition can be used to make functional electrical devices incorporating individual chemical nanostructures. We demonstrate the feasibility here by realizing single-electron tunneling devices incorporating individual 2-nm gold nanocrystals that exhibit Coulomb blockade at low temperature. To fabricate such devices, multiple copper-gold tunnel junctions with a nanometer-sized gap were made using electrodeposition. The electrodes then were rinsed thoroughly in distilled water to remove any residual electrolyte on the substrate, and a dilute aqueous suspension of gold nanocrystals was deposited on them. Transport measurements at T = 4.2 K shows that ~15 % of the devices exhibit Coulomb blockade, while the rest exhibit either single junction tunneling or linear current-voltage characteristics due to tunneling via multiple nanoparticles. Figure 3 illustrates the current-voltage characteristic of a representative device at T = 4.2 K, and illustrates the clear signature of the Coulomb blockade phenomenon.

In summary, we have developed an electrodeposition-based technique to fabricate asymmetric tunnel junctions composed of two different metals. The self-limiting feature of the deposition method in Fig. 2(a) allows the simultaneous fabrication of multiple tunnel junctions. The utility of this technique is demonstrated by realizing single-electron tunneling devices incorporating 2-nm gold nanocrystals. This method enables the fabrication of pairs of metallic electrodes that exhibit distinct magnetic properties or work functions, and should be useful to probe spin polarized tunneling via molecular states[23,26] and to fabricate molecule-sized optoelectronic devices.[24]

**Acknowledgment.** This work is supported by NSF NSEC, US Army, DARPA, and the Packard Foundation. We gratefully acknowledge Alex Champagne and Prof. Daniel Ralph for providing the silicon nitride membranes, and Dr. A.J. Garratt-Reed, Dr. M. Gudiksen, L. Ouyang, and J. J. Urban for their help with the electron microscopy characterizations.

*\* HPark@chemistry.harvard.edu*